\documentclass[final,3p,times]{elsarticle}
\usepackage{}
\usepackage[dvipsnames]{xcolor}

\usepackage{amssymb}

\makeatletter
\newcommand{\rmnum}[1]{\romannumeral #1}
\newcommand{\Rmnum}[1]{\expandafter\@slowromancap\romannumeral #1@}
\makeatother

\journal{Physica A}

\begin{document}

\begin{frontmatter}

{\footnotesize{
\noindent Revised version\\
Original article: {\color{Cyan}Physica A 394 (2014) 166--176}\\
DOI: {\color{Cyan}http://dx.doi.org/10.1016/j.physa.2013.09.054}\\
\\
\\}
}



\title{A complex network analysis of hypertension-related genes}


\author[YNU]{Huan Wang}\ead{whh227@126.com}
\author[YNU]{Chuan-Yun Xu}\ead{kmchyxu@gmail.com}
\author[BJUAS]{Jing-Bo Hu}\ead{joyboble@126.com}
\author[YNU]{Ke-Fei Cao\corref{cor1}}\ead{kfcao163@163.com}

\address[YNU]{Center for Nonlinear Complex Systems, 
Department of Physics, 
School of Physics Science and Technology, 
Yunnan University,\\
Kunming, Yunnan 650091, China}
\address[BJUAS]{Department of Electronic and Electrical Engineering, 
Baoji University of Arts and Sciences, 
Baoji, Shaanxi 721016, China}

\cortext[cor1]{Corresponding author. Tel.: +86 871 65031605.}

\begin{abstract}

In this paper, a network of hypertension-related genes is 
constructed by analyzing the correlations of gene expression 
data among the Dahl salt-sensitive rat and two consomic rat 
strains. The numerical calculations show that this sparse 
and assortative network has small-world and scale-free 
properties. Further, $16$ key hub genes ({\it Col4a1}, {\it Lcn2}, 
{\it Cdk4}, etc.) are determined by introducing an integrated 
centrality and have been confirmed by biological/medical 
research to play important roles in hypertension.
\end{abstract}

\begin{keyword}
Complex network; Hypertension; Integrated centrality; Hub gene
\end{keyword}

\end{frontmatter}



\section{Introduction}
\label{sec:1}

Since the fundamental work on random graphs by Erd\H{o}s and 
R\'{e}nyi \cite{ER1960pmihas}, especially the seminal works on 
the small-world phenomenon by Watts and Strogatz \cite{WS1998n} 
and the scale-free property by Barab\'{a}si and Albert 
\cite{BA1999s}, the study of complex networks has received 
extensive attention. Scientists have found that most real 
networks are neither completely regular nor completely random, 
but usually display a small-world effect and/or a scale-free 
behavior \cite{S2001n,DM2002ap,AB2002rmp,N2003sr}. The research 
results show that complex network theory has been proven to be 
a powerful tool for the analysis in various fields, such as 
the World Wide Web \cite{BKM2000cn,BAJ2000pa}, social networks 
\cite{NP2003pre}, ecological systems \cite{CGA2002prl}, traffic 
systems \cite{vFHH2009epjb,WYY2006pre}, and so on. 

In biological systems, elements that interact or regulate 
each other can be represented by a network, namely, a 
collection of nodes and edges (links) \cite{JTA2000n,%
BO2004nrg,BKS2007febsl,D2011pa,KH2007pa,PSSS2003jtb}. 
At the simplest level, the individual elements are described 
as nodes and their interactions are reduced to edges 
connecting between pairs of nodes. Here, the network-based 
approach is an effective way to discover the collective 
action of the individual parts, namely, the systems-level 
behavior \cite{BKS2007febsl,TSP2009pa,dSS2005jrsi,JHB2008bj}, 
which has been widely used in biochemical and medical research 
and offers a conceptual framework to understand molecular 
mechanisms and disease pathologies.

Hypertension is a major risk factor for cardiovascular 
morbidity and mortality; the worldwide prevalence of 
hypertension is $25\%$ \cite{ML1997l,BCH1995h}. Over the 
years, a great deal of research shows that the arterial 
blood pressure of many essential hypertensive patients 
exhibits an increased sensitivity to dietary salt intake, 
which is known as the salt-sensitive (SS) hypertension 
\cite{WFF2001h,A2002ije}. These people account for about 
$50\%$ of hypertensive patients \cite{RIV2007ndt}. Therefore, 
the studies about SS hypertension contribute to a decrease 
in the level of blood pressure and in the incidence of 
mortality due to cardiovascular complications, and thus 
improve human health \cite{R2000pr,C2006nrg}. Existing 
research for the Dahl SS rat \cite{DHT1962jem,R1982h} and 
consomic rat strains \cite{D1998jh,CRJ2004jp} reveals that 
salt and genetic factors have a major impact on hypertension 
\cite{CRK2001h,JLG2005pg,LLW2008pg}. During the last 30 
years or so, the clinical research and treatment of 
hypertension have improved dramatically 
\cite{CCDN2000c,WMCK2004h,SJJ2011chr}. However, its molecular 
mechanisms and pathologies involved remain intricate and 
elusive.

In the present work, we attempt to study the genes that are 
involved in SS hypertension using the complex network approach. 
We will propose a simple rule to construct the network model 
of hypertension-related genes, where the nodes are individual 
genes and the connections are derived from the expression 
correlations that are based on microarray data. Through 
calculating several statistical indices and analyzing 
topological characteristics of the network, we find out key 
hub genes that play significant roles in hypertension and 
describe their functions. Based on both biological knowledge 
and network theory, this study confirms the role of key genes 
in hypertension from another perspective and provides an 
idea for studying the relation between genes associated with 
hypertension.

This paper is organized as follows. In Section \ref{sec:2} we 
introduce the construction of the network model of 
hypertension-related genes. In Section \ref{sec:3}, we analyze 
the statistical and topological characteristics of the gene 
network, and determine key hub genes by introducing an 
integrated centrality. The biological descriptions of hub 
genes are presented in Section \ref{sec:4}, while Section \ref{sec:5} 
gives concluding remarks.

\section{Construction of network model of hypertension-related genes}
\label{sec:2}

The Dahl SS rat is a widely used genetic model of human 
hypertension, which was proposed by Dahl et al. in the early 
1960s \cite{DHT1962jem,R1982h}. It develops hypertension upon 
exposure to a high-salt intake. The consomic rat strains, used 
as the normotensive control for the Dahl SS rat, are generated 
by substituting a chromosome or a part of a chromosome from a 
normal rat strain for the corresponding genomic region of the 
SS rat \cite{D1998jh,CRJ2004jp,LLW2008pg}. Previous research 
showed that substitution of chromosome 13 or 18 could 
significantly attenuate hypertension \cite{CRK2001h,JLG2005pg}. 
The two consomic rat strains (generated by substituting 
chromosomes 13 and 18, respectively) have genetic homology with 
the Dahl SS rat, but lead to the amelioration of hypertension; 
and the segregation of hypertensive phenotype is regarded as 
the result of different gene expression patterns among the 
three rat strains caused by substituting chromosomes. Therefore, 
our study focuses on the analysis of gene expression data (GED) 
among the Dahl SS rat and two consomic rat strains: 
$H$ (GED for SS rat with high blood pressure), 
$S^{\mathrm{\Rmnum{1}}}$ (GED for substitution of chromosome 13) and 
$S^{\mathrm{\Rmnum{2}}}$ (GED for substitution of chromosome 18), 
where all three strains of rats were maintained on a high 
sodium intake for two weeks and gene expression profiles were 
examined using microarrays (see http://pga.mcw.edu). All data of 
$H$, $S^{\mathrm{\Rmnum{1}}}$ and $S^{\mathrm{\Rmnum{2}}}$ are given in 
three columns (SS2wk, SS13 2wk, and SS18 2wk) in Table S2 of 
Supplemental Figures and Tables of Ref.~\cite{LLW2008pg}. In 
Table \ref{dataformat} we list detailed data of some randomly 
selected genes and feature genes calculated in Section \ref{sec:3}.

Let us consider the network $G_H=(V_H,E_H)$, where $V_H=\{v_i\}$ 
$(i=1,2,\ldots,N)$ is the set of $N$ nodes, and $E_H=\{v_i,v_j\}$ 
the set of edges or connections between nodes. We will use the 
following notation: $A_{ij}=1$ indicates that there is an edge 
between nodes $v_i$ and $v_j$; and $A_{ij}=0$ otherwise. Our 
gene network model is constructed from the correlations based on 
GED in three rat strains in two steps, which is designed to 
explore the relationship between genetic change trend and 
hypertensive phenotype.

{\it Step 1. Calculation of change ratios of GED:} The $N=335$ 
hypertension-related genes are served as nodes of the network, 
each node represents an individual gene shown as gene symbol 
or CloneID (for the gene without a gene symbol). For each of 
the $335$ nodes, we define the change ratios 
$R^{\mathrm{\Rmnum{1}}}_{i}$ and $R^{\mathrm{\Rmnum{2}}}_{i}$ of GED 
between $\{ S^{\mathrm{\Rmnum{1}}}_{i},S^{\mathrm{\Rmnum{2}}}_{i}\}$ and 
$H_{i}$ as follows: 
\begin{equation}
\left \{ \begin{array}{l}
        R^{\mathrm{\Rmnum{1}}}_{i}=(S^{\mathrm{\Rmnum{1}}}_{i}-H_{i})/|H_{i}|,\\
        R^{\mathrm{\Rmnum{2}}}_{i}=(S^{\mathrm{\Rmnum{2}}}_{i}-H_{i})/|H_{i}|,\\
        \end{array} \right.
\end{equation}
where $i=1,2,\ldots,335$. Then we can calculate the change ratios 
$R^{\mathrm{\Rmnum{1}}}_{i}$ and $R^{\mathrm{\Rmnum{2}}}_{i}$ separately 
(these data are shown as $R^{\mathrm{\Rmnum{1}}}$ and $R^{\mathrm{\Rmnum{2}}}$ 
in Table \ref{dataformat}). It is easy to obtain the averages of 
two groups of change ratios $\{R^{\mathrm{\Rmnum{1}}}_{i}\}$ and 
$\{R^{\mathrm{\Rmnum{2}}}_{i}\}$, respectively: 
$T^{\mathrm{\Rmnum{1}}}=0.150725$, 
and $T^{\mathrm{\Rmnum{2}}}=0.141180$; which can be used as thresholds 
to determine whether a connection should be made between two nodes 
in the next step. It should be indicated that there are many ways 
of choosing the threshold, a higher threshold would result in a 
higher average degree and a greater influence or disturbance among 
nodes (genes), whereas a lower threshold would result in fewer 
connections but a greater possibility of the actual correlation of 
genes. So the average change ratio would be a reasonable choice of 
the threshold to make reliable connections and a moderate scale of 
the network. Indeed, setting the threshold a little higher or lower 
than the average would not disturb the clustered (not paired) 
relationship of genes considered in this paper.

\begin{table}[t]
\centering\scriptsize
\caption{Some GED ($H$, $S^{\mathrm{\Rmnum{1}}}$ and $S^{\mathrm{\Rmnum{2}}}$) of 
three rat strains and change ratios ($R^{\mathrm{\Rmnum{1}}}$ and 
$R^{\mathrm{\Rmnum{2}}}$) of two groups. In this paper, we use 
``{\it Hist1h2ai}'' as the abbreviation of gene symbol 
``{\it Hist1h2ai\_predicted /// Hist1h4a\_predicted}''.}
\label{dataformat}
\begin{tabular}[c]{lrrrrr}
\hline
Gene & $H$\quad\,\,\,\ \ \ \ \ \ \ \ \ \ 
&  $S^{\mathrm{\Rmnum{1}}}$\quad\,\,\,\ \ \ \ \ \ \ \ \ 
&  $S^{\mathrm{\Rmnum{2}}}$\quad\,\,\,\ \ \ \ \ \ \ \ 
&  $R^{\mathrm{\Rmnum{1}}}$\quad\,\,\,\ \ \ \ \ \ \ \ \ 
&  $R^{\mathrm{\Rmnum{2}}}$\quad\,\,\,\ \ \ \ \ \ \ \ \\
\hline
{\it Ssg1}       & 0.760115 & 0.039832 & 0.084195 & $-$0.947598 & $-$0.889234\\
{\it Lcn2}       & 0.915950 & 0.089252 & 0.114858 & $-$0.902558 & $-$0.874602\\
{\it Ociad1}     & $-$0.410164 & 0.075833 & 0.108042 & 1.184886 & 1.263413 \\
{\it Nr1d2}      & 0.503251 & 0.036939 & 0.056712 & $-$0.926599 & $-$0.887308\\
{\it Tagln}      & 0.765658 & 0.130215 & 0.223917 & $-$0.829931 & $-$0.707550\\
{\it Fzd2}       & 0.474655 & 0.086283 & 0.068861 & $-$0.818220 & $-$0.854925\\
{\it RGICL83}    & 0.456126 & $-$0.024038 & 0.072557 & $-$1.052701 & $-$0.840928\\
{\it Cdc2a}      & 0.583242 & 0.106746 & 0.118771 & $-$0.816978 & $-$0.796360\\
{\it Colec12}    & 0.450878 & 0.117010 & 0.230425 & $-$0.740483 & $-$0.488942\\
{\it RGIEJ34}    & 0.628989 & 0.075356 & 0.140478 & $-$0.880196 & $-$0.776660\\
{\it RGIHD68}    & 0.608212 & 0.213125 & 0.258199 & $-$0.649589 & $-$0.575479\\
{\it Mcm6}       & 0.255819 & 0.112129 & 0.061829 & $-$0.561684 & $-$0.758311\\
{\it Ctsd}       & 0.606232 & 0.088551 & 0.151729 & $-$0.853932 & $-$0.749719\\
{\it Casp6}  & 0.501480 & 0.101879 & 0.095765 & $-$0.796843 & $-$0.809036\\
{\it Timp1}      & 0.589334 & 0.103506 & 0.122895 & $-$0.824368 & $-$0.791467\\
{\it Ctsl}       & 0.277923 & $-$0.046910 & 0.023243 & $-$1.168789 & $-$0.916368\\
{\it MCWA09\_96} & 0.372041 & 0.049722 & 0.087101 & $-$0.866352 & $-$0.765883\\
{\it Nudt4}      &$-$0.368880 & $-$0.142909 & $-$0.061423 & 0.612586 & 0.833487\\
{\it Hist1h2ai}  & 0.532325 & 0.100425 & 0.099606 & $-$0.811346 & $-$0.812884\\
{\it MCWA13\_68} & 0.432117 & 0.049337 & 0.096514 & $-$0.885824 & $-$0.776649\\
{\it Hfe}        & 0.339892 & 0.069570 & 0.055555 & $-$0.795316 & $-$0.836552\\
{\it MCWA10\_48} & 0.488626 & 0.061358 & 0.087510 & $-$0.874427 & $-$0.820907\\
{\it Rbp1}       & 0.487553 & 0.019423 & 0.108073 & $-$0.960162 & $-$0.778336\\
{\it Cdk4}       & 0.297970 & 0.043814 & 0.051296 & $-$0.852959 & $-$0.827847\\
{\it Sdc1}       & 0.447238 & 0.047824 & 0.105806 & $-$0.893068 & $-$0.763424\\
{\it Fbn1}       & 0.290349 & 0.047677 & 0.075306 & $-$0.835793 & $-$0.740637\\
{\it Gpnmb}      & 0.374620 & $-$0.004214 & 0.052804 & $-$1.011249 & $-$0.859047\\
{\it MCWA11\_14} & 0.260709 & 0.063897 & 0.033333 & $-$0.754912 & $-$0.872143\\
{\it MCW075\_23} & 0.301193 & $-$0.020668 & 0.069398 & $-$1.068621 & $-$0.769590\\
{\it MCW067\_02} & 0.297496 & $-$0.081443 & 0.058038 & $-$1.273761 & $-$0.804913\\
{\it Slc25a10}   & 0.268964 & 0.073774 & 0.045496 & $-$0.725709 & $-$0.830849\\
{\it Usp48}      & 0.166792 & 0.011649 & 0.039235 & $-$0.930159 & $-$0.764767\\
{\it Col4a1}     & 0.258305 & 0.043254 & 0.038870 & $-$0.832545 & $-$0.849517\\
{\it B2m}        & 0.127737 & 0.006545 & 0.029585 & $-$0.948760 & $-$0.768391\\
{\it Shc1}       & 0.257072 & 0.036985 & 0.053275 & $-$0.856130 & $-$0.792763\\
{\it Fstl1}      & 0.177793 & 0.025137 & 0.025407 & $-$0.858616 & $-$0.857100\\
{\it Rgs2}       & 0.174529 & 0.115578 & $-$0.027143 & $-$0.337772 & $-$1.155523\\
\hline
\end{tabular}
\end{table}

\begin{figure}[t]
\begin{center}
\includegraphics[width=13.8cm]{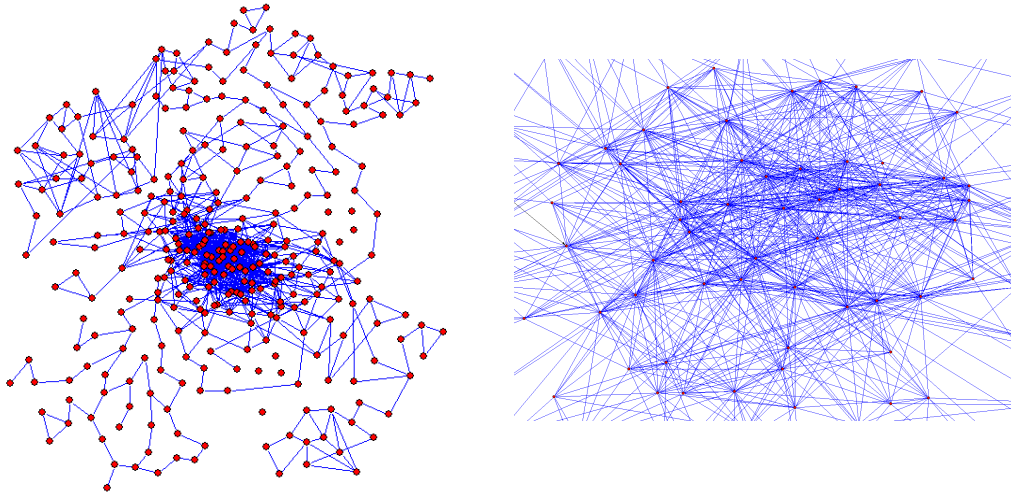}
\caption{Illustration of the network of hypertension-related genes 
for the SS rat with all $335$ nodes (left) and a magnification of 
its central part (right).}
\label{gene network}
\end{center}
\end{figure}

{\it Step 2. Establishment of connections:} Chromosome substitution 
could affect genes located on the substituted chromosomes and other 
related chromosomes, which would up-regulate or down-regulate the 
expression levels of targeted genes, and unrelated genes would keep 
their expression levels unchanged. Correspondingly, the change ratios 
of unrelated genes would tend to zero. Therefore, the change ratios 
could reflect the relationship among genes. We consider the 
correlations between genes by contrasting the change ratios between 
any two genes. Specifically, we can compare the change ratios 
$R^{\mathrm{\Rmnum{1}}}_{i}$ and $R^{\mathrm{\Rmnum{2}}}_{i}$ among all $335$ 
nodes. If the trend of changes between two nodes is similar in both 
$R^{\mathrm{\Rmnum{1}}}$ and $R^{\mathrm{\Rmnum{2}}}$ according to 
(\ref{adjacentmatrix}), then a connection is made between such two 
nodes (genes): 
\begin{equation}
A_{ij}=\left \{ \begin{array}{lcl}
        1 & & \mbox{if}\  |R^{\mathrm{\Rmnum{1}}}_{i}-R^{\mathrm{\Rmnum{1}}}_{j}|
        \leq T^{\mathrm{\Rmnum{1}}} \ \mbox{and}\  
        |R^{\mathrm{\Rmnum{2}}}_{i}-R^{\mathrm{\Rmnum{2}}}_{j}|\leq T^{\mathrm{\Rmnum{2}}}; \\
        0 & & \mbox{if}\  |R^{\mathrm{\Rmnum{1}}}_{i}-R^{\mathrm{\Rmnum{1}}}_{j}|
        > T^{\mathrm{\Rmnum{1}}}  \ \mbox{or}\  
        |R^{\mathrm{\Rmnum{2}}}_{i}-R^{\mathrm{\Rmnum{2}}}_{j}|> T^{\mathrm{\Rmnum{2}}}. 
        \label{adjacentmatrix}
        \end{array} \right.
\end{equation}
Here, $i,j=1,2,\ldots,335$ and $i\neq j$. $A_{ij}=1$ indicates that 
substitutions of chromosomes 13 and 18 have consistent and similar 
effects on the genes, while both substitutions could attenuate 
hypertension; consequently, there is a certain correlation between 
two genes $i$ and $j$. In such a way, we have constructed the 
network of hypertension-related genes, which contains $335$ nodes 
(genes) and $1280$ edges (connections or links). In 
Fig.~\ref{gene network}, we show the schematic diagram of the gene 
network for the SS rat with all $335$ nodes and an enlarged view of 
its central part.

\section{Statistical and topological characteristics of gene network}
\label{sec:3}

To analyze the network of hypertension-related genes, we focus on 
the following indices: degree distributions, sparsity, average path 
length, clustering coefficient, assortativity and three centrality 
indices (degree centrality, betweenness centrality and closeness 
centrality). Further, we will introduce an integrated centrality 
to determine hub genes in the network.

\subsection{Degree distribution}
\label{subsec:3.1}

\begin{figure}[!]
\begin{center}
\includegraphics[width=12.2cm]{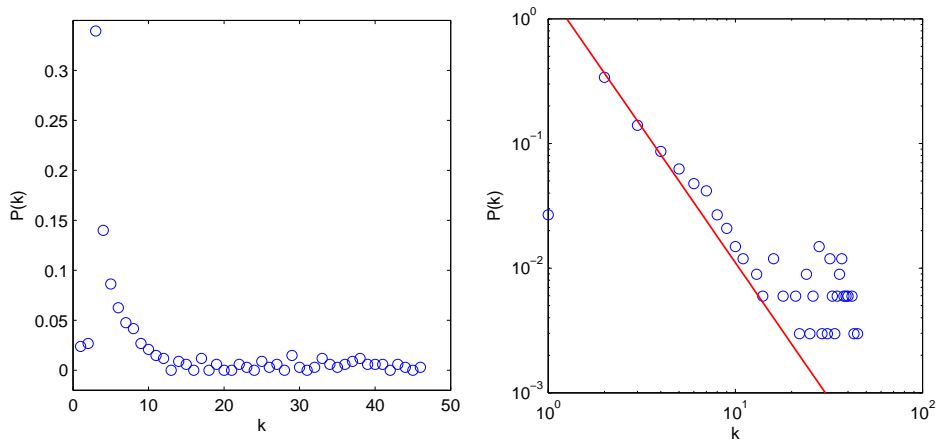}
\caption{The degree distributions for the network of 
hypertension-related genes in two coordinates: $k$--$P(k)$ (left) and 
log--log degree distribution in logarithmic coordinates (right). In 
this figure, $P(k)$ denotes the probability of a node having degree 
$k$. The power-law exponent $\gamma$ is $2.173$.}
\label{degree distribution}
\end{center}
\end{figure}

The most elementary characteristic of a node is its degree, denoted 
by $k$, which tells us the number of links (connections or edges) the 
node has to other nodes. The degree $k_i$ of a node $i$ is the total 
number of its links. The average of $k_i$ over all nodes is called 
the average degree of the network, and is denoted as 
$\langle k\rangle $. The spread in node degree is characterized by 
a distribution function $P(k)$, which can quantify the diversity of 
the whole network. The degree distribution $P(k)$ gives the 
probability that a randomly selected node has exactly $k$ links 
(edges). $P(k)$ is obtained by counting the number $N(k)$ of nodes 
that have $k=1,2,\ldots$ links and dividing by the total number $N$ 
of nodes \cite{DM2002ap,AB2002rmp,N2003sr}. The degree distribution 
is one of the most important statistical characteristics of networks. 

Fig.~\ref{degree distribution} plots the degree distributions for 
the nodes of the gene network in two coordinates. The illustrations 
show that the probability that a gene can link with $k$ other genes 
decays as a power-law of the form $P(k)\sim k^{-\gamma}$, where the 
exponent $\gamma$ is about $2.173$ suggesting that the gene network 
has a scale-free topology. A distinguishing feature of such a 
scale-free network is the existence of a small number of highly 
connected nodes, known as hubs, which often determine the network 
properties and are more important than a large number of other less 
connected nodes. These hubs correspond to key (feature) genes which 
play important roles in hypertension.

\subsection{Sparsity}
\label{subsec:3.2}

The gene network involves $335$ nodes. If it is fully connected, then 
the total number of edges of the network is $C^{2}_{335}=55945$. 
However, the number of edges in our gene network is only $1280$, so 
the ratio between them is $2.288\%$. From another point of view, 
the average degree (or average number of connections) 
$\langle k \rangle$ is about $7.642$. Thus it can be seen from both 
the ratio and the average degree $\langle k \rangle$ that the gene 
network is sparse.

\subsection{Average path length and clustering coefficient}
\label{subsec:3.3}

We now briefly recall some basic notions of complex networks. Two 
nodes of a network are connected if a path, namely a sequence of 
adjacent nodes, links them. There are many alternative paths between 
two nodes, the path with the smallest number of links (edges) between 
the selected nodes is called the shortest path. The distance $d_{ij}$ 
between two nodes $i$ and $j$ is defined as the number of edges along 
the shortest path connecting them. The diameter $D$ is the maximum 
distance between any pair of nodes in the network, i.e. 
\begin{eqnarray}
D=\max\{d_{ij}\}.
\end{eqnarray}
The average path length (also called characteristic path length) $L$ 
is defined as the mean distance between two nodes, averaged over all 
pairs of nodes, i.e. 
\begin{eqnarray}
L=\frac{1}{N(N-1)}\sum_{i\neq j} d_{ij},
\end{eqnarray}
here $L$ determines the effective ``size'' of a network, and offers 
a measure of the overall navigability of a network \cite{X2003tag}.

A clustering coefficient can be defined to describe the cohesiveness 
of the neighborhood of a node \cite{WS1998n,N2003sr}. In a network, 
suppose that a node $i$ has $k_{i}$ edges, the $k_{i}$ nodes are 
the neighbors of node $i$. The clustering coefficient $c_{i}$ is 
defined as the ratio between the number $e_{i}$ of edges that 
actually link the $k_{i}$ neighbors of node $i$ to each other and 
the total possible number of edges among them, i.e. 
\begin{eqnarray}
c_{i}=\frac{2e_{i}}{k_{i}(k_{i}-1)} & {\mathrm{for\ }}k_{i} \geq 2.
\label{clustering}
\end{eqnarray}
The clustering coefficient $C$ of the whole network is the average 
of $c_{i}$ over all $i$, which characterizes the overall tendency 
of nodes to form clusters, clearly, $C\leq 1$.

The small-world effect consists of two properties: a short average 
path length and a relatively high clustering coefficient. In the 
following, we will give the calculation results about these two 
properties and show that the gene network is a small-world network.

The gene network exhibits a very short average path length: $L$ 
is about $5.428$, and proportional to the logarithm of the 
network size $N$, i.e., $L \sim \log(N)$. On the other hand, 
the maximum distance (i.e., the diameter) $D$ is only $12$. That 
is, at most twelve hops separate any two genes in the $1280$ links 
of the gene network.

For the gene network, the clustering coefficient is calculated to 
be $C=0.6063$ according to (\ref{clustering}). Compared with the 
clustering coefficient $C_{\mathrm ER}=\langle k \rangle/N=0.0228$ 
of a corresponding Erd\H{o}s--R\'{e}nyi random graph, the clustering 
coefficient $C$ of the gene network is about $27$ times higher than 
that of the random graph. From what is discussed above, we can draw 
a conclusion that the gene network has the small-world property 
characterized by small $L$ and large $C$.

\subsection{Assortativity}
\label{subsec:3.4}

To describe degree correlations between neighboring nodes in a 
network, the concept of assortativity is introduced \cite{N2002prl}. 
A network is said to be assortative if the nodes with many connections 
tend to be connected to other nodes with many connections; otherwise, 
it is said to be disassortative if the nodes with many connections 
tend to be connected to other nodes with few connections. Most social 
networks usually exhibit assortativity \cite{NP2003pre}. The 
assortativity can be described by measuring the correlation between 
the degrees of neighboring nodes in terms of the mean Pearson 
correlation coefficient. For any link $i$, let $x_{i}$ and $y_{i}$ 
be the degrees of the two vertices connected by the $i$th edge, with 
$i=1,\ldots,E$ ($E$ is the number of edges in the network), then the 
assortativity coefficient of the network is given by \cite{N2002prl}: 
\begin{eqnarray}
r=\frac{E^{-1}\sum\limits_{i}x_{i}y_{i}
  -\left[E^{-1}\sum\limits_{i}\frac{1}{2}(x_{i}+y_{i})\right]^{2}}
  {E^{-1}\sum\limits_{i}\frac{1}{2}(x_{i}^{2}+y_{i}^{2})
  -\left[E^{-1}\sum\limits_{i}\frac{1}{2}(x_{i}+y_{i})\right]^{2}}.
\label{ass}
\end{eqnarray}
The network is assortative if $r>0$; otherwise it is disassortative 
if $r<0$.

For the network of hypertension-related genes, the assortativity 
coefficient is calculated to be $r=0.4065$ from (\ref{ass}). Based 
on this value, we can say that this gene network exhibits an 
assortative behavior like most social networks, but unlike most 
biological networks.

\subsection{Centrality}
\label{subsec:3.5}

\subsubsection{Three centrality indices}
\label{subsec:3.5.1}

\begin{table}[!tbp]
\centering\scriptsize
\caption{Top $20$ values of degree centrality $C_{d}$, betweenness 
centrality $C_{b}$, closeness centrality $C_{c}$, and integrated 
centrality $C_{\mathrm{intgr}}$ of hypertension-related genes.}
\label{datacentrality}
\begin{tabular}[c]{llp{1mm}llp{1mm}llp{1mm}ll}
\hline
Gene             & $C_{d}$ &  & Gene             & $C_{b}$ &  & Gene             & $C_{c}$ &  & 
Gene             & $C_{\mathrm{intgr}}$\\
\cline{1-2}\cline{4-5}\cline{7-8}\cline{10-11}
{\it Lcn2}       & 0.1347  &  & {\it Cdk4}       & 0.06477  &  & {\it Col4a1}     & 0.1771  &  & 
{\it Col4a1}     & 0.9591\\
{\it Ctsd}       & 0.1287  &  & {\it Col4a1}     & 0.06114  &  & {\it Cdk4}       & 0.1755  &  & 
{\it Lcn2}       & 0.9560\\
{\it Col4a1}     & 0.1257  &  & {\it Usp48}      & 0.05826  &  & {\it Usp48}      & 0.1752  &  & 
{\it Cdk4}       & 0.9525\\
{\it Fstl1}      & 0.1257  &  & {\it Lcn2}       & 0.05700  &  & {\it Nr1d2}      & 0.1752  &  & 
{\it Fstl1}      & 0.8741\\
{\it Shc1}       & 0.1198  &  & {\it Fstl1}      & 0.04612  &  & {\it Lcn2}       & 0.1750  &  & 
{\it Usp48}      & 0.8666\\
{\it Sdc1}       & 0.1198  &  & {\it Shc1}       & 0.04538  &  & {\it Shc1}       & 0.1746  &  & 
{\it Shc1}       & 0.8584\\
{\it Cdk4}       & 0.1168  &  & {\it Nr1d2}      & 0.04136  &  & {\it Cdc2a}      & 0.1746  &  & 
{\it Nr1d2}      & 0.8241\\
{\it MCWA09\_96} & 0.1168  &  & {\it Fzd2}       & 0.03788  &  & {\it Fzd2}       & 0.1743  &  & 
{\it Fzd2}       & 0.8044\\
{\it Nr1d2}      & 0.1138  &  & {\it Cdc2a}      & 0.03555  &  & {\it Hist1h2ai}  & 0.1734  &  & 
{\it Cdc2a}      & 0.7856\\
{\it Fzd2}       & 0.1138  &  & {\it Hist1h2ai}  & 0.03499  &  & {\it Casp6}      & 0.1734  &  & 
{\it Hist1h2ai}  & 0.7730\\
{\it Cdc2a}      & 0.1108  &  & {\it RGICL83}    & 0.02879  &  & {\it Fstl1}      & 0.1730  &  & 
{\it Ctsd}       & 0.7722\\
{\it Fbn1}       & 0.1108  &  & {\it Ctsd}       & 0.02488  &  & {\it Ctsd}       & 0.1730  &  & 
{\it Casp6}      & 0.7191\\
{\it Timp1}      & 0.1108  &  & {\it Casp6}      & 0.02453  &  & {\it Timp1}      & 0.1730  &  & 
{\it Timp1}      & 0.7183\\
{\it Hfe}        & 0.1108  &  & {\it Timp1}      & 0.02304  &  & {\it Ssg1}       & 0.1730  &  & 
{\it Fbn1}       & 0.7052\\
{\it Hist1h2ai}  & 0.1078  &  & {\it Fbn1}       & 0.02127  &  & {\it MCWA09\_96} & 0.1712  &  & 
{\it MCWA09\_96} & 0.7043\\
{\it Casp6}      & 0.1078  &  & {\it MCW067\_02} & 0.02031  &  & {\it Fbn1}       & 0.1709  &  & 
{\it Sdc1}       & 0.6925\\
{\it MCWA13\_68} & 0.1078  &  & {\it MCWA09\_96} & 0.01812  &  & {\it Sdc1}       & 0.1706  &  & 
{\it Ssg1}       & 0.6263\\
{\it RGIEJ34}    & 0.1048  &  & {\it RGIHD68}    & 0.01539  &  & {\it MCWA10\_48} & 0.1697  &  & 
{\it MCWA10\_48} & 0.6260\\
{\it MCWA11\_14} & 0.1048  &  & {\it Sdc1}       & 0.01460  &  & {\it B2m}        & 0.1691  &  & 
{\it Hfe}        & 0.6236\\
{\it B2m}        & 0.1018  &  & {\it MCW075\_23} & 0.01445  &  & {\it Hfe}        & 0.1691  &  & 
{\it RGIEJ34}    & 0.6148\\
\hline
\end{tabular}
\end{table}

The study of centrality aims at finding out the centralization nodes 
in the network. There are three centrality indices widely used in 
network analysis: degree centrality, betweenness centrality, and 
closeness centrality \cite{B1965bs,F1979sn}. These centrality 
indices determine the relative importance of a node in the network.

The degree centrality of a given node $i$ is the proportion of other 
nodes that are adjacent to $i$ \cite{F1979sn}, i.e. 
\begin{eqnarray}
C_{d}(i)=\frac{k_{i}}{N-1},
\label{degreecentrality}
\end{eqnarray}
here $k_{i}$ is the degree of node $i$, and $N-1$ the maximum possible 
degree of the network. $C_{d}(i)$ is a structural measure of node 
centrality based on the degree of node $i$.

The second index of node centrality is called betweenness centrality. 
This index is based upon the frequency with which a node falls between 
pairs of other nodes on the shortest or geodesic paths connecting them. 
In a network of $N$ nodes, a geodesic is the shortest path between two 
nodes; the betweenness centrality of a node $i$ is defined as the 
proportion of all geodesics between pairs of other nodes that include 
this node $i$ \cite{F1979sn}: 
\begin{eqnarray}
C_{b}(i)=\sum_{j(<k)}^{N}\sum_{k}^{N}\frac{g_{jk}(i)}{g_{jk}},
\label{beteencentrality}
\end{eqnarray}
where $g_{jk}$ is the number of geodesics connecting nodes $j$ and $k$, 
and $g_{jk}(i)$ the number of geodesics connecting the two nodes $j$ 
and $k$ that contain node $i$.

With the concept of distance described in Section \ref{subsec:3.3}, 
the third index of centrality is defined, which is called closeness 
centrality. The closeness centrality of a node $i$ is the number of 
other nodes divided by the sum of the distances between the node $i$ 
and all others \cite{B1965bs,F1979sn}: 
\begin{eqnarray}
C_{c}(i)=(L_{i})^{-1}=\frac{N-1}{\sum\limits_{j=1}^{N}d_{ij}},
\label{closecentrality}
\end{eqnarray}
here, $L_{i}$ is the average distance between the node $i$ and all 
other nodes, and $d_{ij}$ the distance between nodes $i$ and $j$. 
For the network of hypertension-related genes, the three centrality 
indices are calculated and listed in Table \ref{datacentrality}, 
respectively (we only show the top $20$ values of each centrality 
index).

\begin{figure}[!]
\begin{center}
\includegraphics[width=12.3cm]{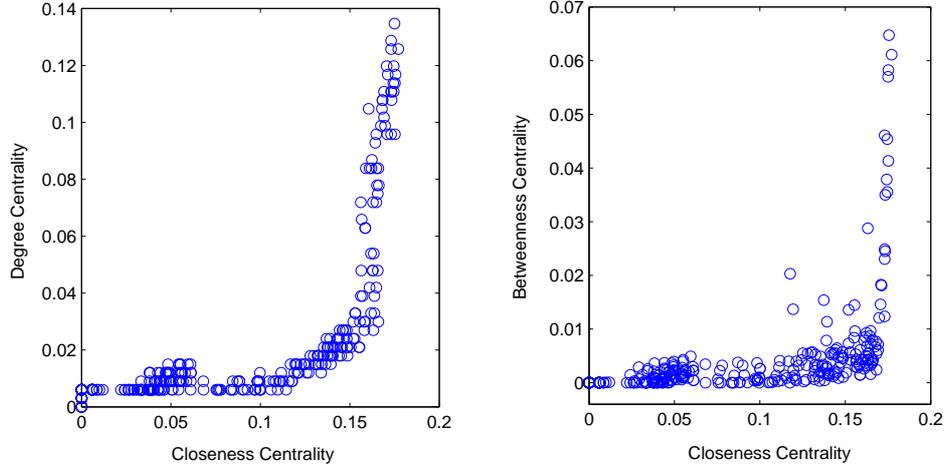}
\caption{Correspondence among degree centrality $C_{d}$, betweenness 
centrality $C_{b}$, and closeness centrality $C_{c}$ of 
hypertension-related genes: $C_{d}$ versus $C_{c}$ (left), and 
$C_{b}$ versus $C_{c}$ (right).}
\label{bcd-centrality}
\end{center}
\end{figure}

The degree centrality can be interpreted in terms of the immediate 
influence of a node in the network, and the betweenness centrality 
is a measurement of the shortest paths including the node. The 
closeness centrality depicts how close a node is to all other nodes, 
and reflects the ability of influence of a node on other nodes 
through the network. Fig.~\ref{bcd-centrality} describes the 
correspondence among degree centrality $C_{d}$, betweenness 
centrality $C_{b}$, and closeness centrality $C_{c}$ of 
hypertension-related genes. The illustration shows that a small 
number of nodes with high degree and betweenness centrality have 
almost the same high closeness centrality; while most nodes have 
low degree and betweenness centrality, whose closeness centrality 
distributes widely.

\subsubsection{Determination of hub genes}
\label{subsec:3.5.2}

As calculated above, the network of hypertension-related genes is 
scale-free, which shows that the network's properties are often 
determined by a small number of highly connected nodes (i.e., hubs). 
So the determination of hubs is an important issue. We know that 
the centrality measurements can accurately and quickly find out 
relatively important nodes (hubs) in complex networks. In general, 
one can recognize hubs according to any one of three centrality 
indices $C_{d}$, $C_{b}$ and $C_{c}$. However, in order to 
fully reflect the contribution of all these three centrality 
indices, here we will determine key ``hub'' genes by 
considerations from the following two aspects.

(\rmnum{1}) {\it Direct consideration from three high centrality indices}

\begin{figure}[!]
\begin{center}
\includegraphics[width=8.55cm]{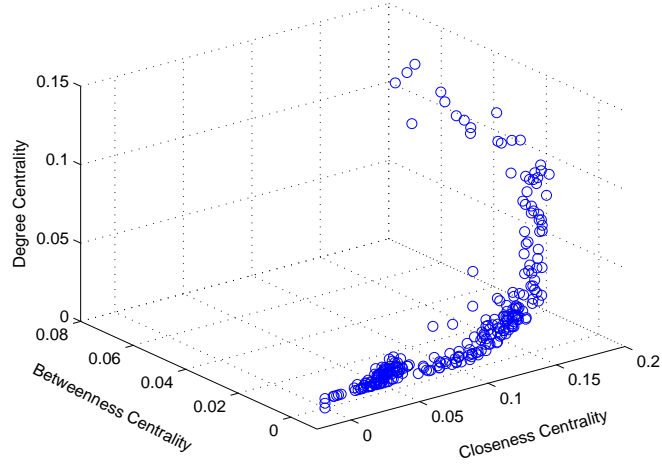}
\caption{Three-dimensional diagram of degree centrality $C_{d}$, 
betweenness centrality $C_{b}$, and closeness centrality $C_{c}$ 
of hypertension-related genes.}
\label{3D diagram}
\end{center}
\end{figure}

We can explore node centrality in the gene network through directly 
calculating the three centrality indices and analyzing the 
relationship among them. The betweenness and closeness centrality 
are based on the shortest path, indicating that local perturbations 
to hub genes could spread to the whole network very rapidly. 
Fig.~\ref{3D diagram} illustrates the three-dimensional diagram of 
degree centrality, betweenness centrality and closeness centrality 
of the gene network. It can be seen from Fig.~\ref{3D diagram} that a 
small number of nodes (hubs) have high values of three centrality 
indices.

\begin{figure}[!]
\begin{center}
\includegraphics[width=9.5cm]{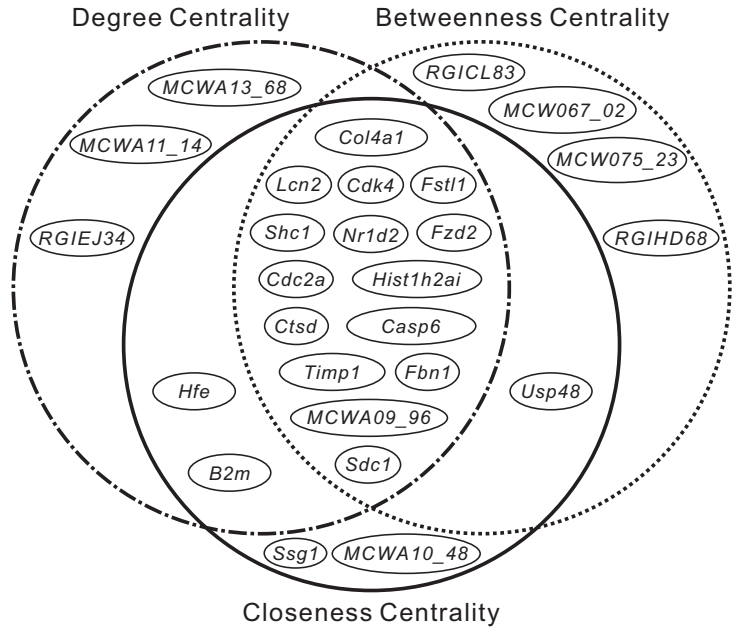}
\caption{Distribution diagram of some important hypertension-related 
genes based on analysis of three centrality indices $C_{d}$, $C_{b}$ 
and $C_{c}$.}
\label{gene centrality}
\end{center}
\end{figure}

\begin{figure}[!]
\begin{center}
\includegraphics[width=6.5cm]{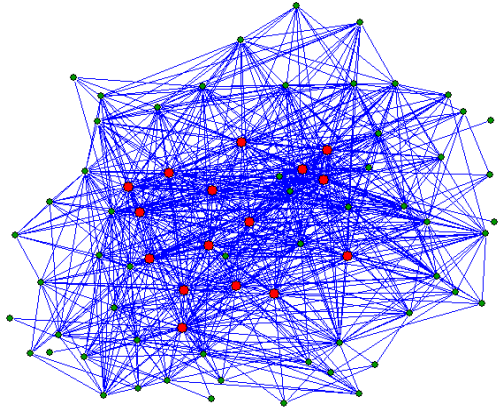}
\caption{Illustration of all connections of $16$ hub genes (big 
red dots) with other genes (small green dots).} 
\label{hubs}
\end{center}
\end{figure}

Fig.~\ref{gene centrality} illustrates the distribution of certain 
genes according to the values of three centrality indices 
$\{C_{d},C_{b},C_{c}\}$ in Table \ref{datacentrality}, where each 
circle shows the top $20$ genes of each centrality index. Thus the 
intersection of three circles directly gives $15$ important hub 
genes in the network of hypertension-related genes.

(\rmnum{2}) {\it Quantitative consideration from integrated centrality}

We can also quantitatively determine hub genes by introducing an 
integrated centrality index. To this end, we first convert (rescale) 
three centrality indices into three relative centrality indices: 
$C_{d}(i)/C_{d,\max}$, $C_{b}(i)/C_{b,\max}$ and 
$C_{c}(i)/C_{c,\max}$, where $C_{d,\max}$, $C_{b,\max}$ 
and $C_{c,\max}$ are the maximums of $\{C_{d}\}$, $\{C_{b}\}$ 
and $\{C_{c}\}$, respectively. Then we can introduce an integrated 
centrality of node $i$, defined as the average of them: 
\begin{eqnarray}
C_{\mathrm{intgr}}(i)=\frac{1}{3}\left[ C_{d}(i)/C_{d,\max}+%
C_{b}(i)/C_{b,\max}+C_{c}(i)/C_{c,\max} \right].
\label{compcentrality}
\end{eqnarray}
Obviously, $C_{\mathrm{intgr}}(i)$ has a value between $0$ and $1$. For 
different nodes (genes), the values of $C_{\mathrm{intgr}}$ can be easily 
calculated, and the top $20$ values of $C_{\mathrm{intgr}}$ are listed 
in Table \ref{datacentrality}. We observe that the first three genes 
of the top $3$ $C_{\mathrm{intgr}}$ values correspond to the maximums of 
the three centrality indices, respectively: 
{\it Col4a1} ($C_{c,\max}=0.1771$), 
{\it Lcn2} ($C_{d,\max}=0.1347$), 
and {\it Cdk4} ($C_{b,\max}=0.06477$). We can see that for the top 
$16$ genes of $C_{\mathrm{intgr}}>0.69$, except for the gene {\it Usp48}, 
all other $15$ genes are consistent with hub genes determined in (i). 
Although not included in hub genes in (\rmnum{1}) (for the reason 
that the degree centrality does not enter the top $20$ values), {\it Usp48} 
has a high value of $C_{\mathrm{intgr}}$ due to its high closeness 
and betweenness centrality, so we should include it in hub genes.

Thus, combining with considerations (\rmnum{1}) and (\rmnum{2}), 
we can determine $16$ important hub genes in the network of 
hypertension-related genes: {\it Col4a1}, {\it Lcn2}, {\it Cdk4}, 
{\it Fstl1}, {\it Usp48}, {\it Shc1}, {\it Nr1d2}, {\it Fzd2}, 
{\it Cdc2a}, {\it Hist1h2ai}, {\it Ctsd}, {\it Casp6}, {\it Timp1}, 
{\it Fbn1}, {\it MCWA09\_96}, and {\it Sdc1}. Fig.~\ref{hubs} shows all 
connections of these $16$ hub genes with other genes. These hub 
genes have the integrated centrality of $C_{\mathrm{intgr}}>0.69$, 
which is about $70\%$ of its maximum ($0.9591$); however, it 
should be indicated that there is no strict significance 
threshold for $C_{\mathrm{intgr}}$, and one can also lower the 
threshold to enable more genes to be included in hub genes.

Moreover, as shown in Figs. \ref{bcd-centrality} and 
\ref{3D diagram}, we can observe that most nodes with small 
degree have widely distributed closeness centrality, indicating 
that despite of low value of degree, a node could have major 
impact in the network due to high closeness centrality. 
Therefore, for the above two considerations, since 
$C_{\mathrm{intgr}}$ can comprehensively and quantitatively 
reflect the contribution of three centrality indices 
$C_{d}$, $C_{b}$ and $C_{c}$, we suggest to determine 
hub genes by using $C_{\mathrm{intgr}}$ (i.e., consideration 
(\rmnum{2})) for simplicity.

\section{Biological descriptions of hub genes}
\label{sec:4}

In the gene network, hubs typically correspond to key 
genes and are closely related to hypertension. In Table 
\ref{genefunction} we list the official full names, gene IDs, 
and biological functions of $16$ hub genes (cf. Refs. 
\cite{TSS2009ccg,vABS2010hmg,OTC2011ceh,DEMT2001h,LWX2011art,%
AVW2011c,NYM2010cc,LBM2013gn,GDA2002a,vGDS2002cr,YQG2006bbrc,%
CWH2010jmcc,T1992sjcli,TGK2009k,BZL2005p,TNB2004ajh,%
SLW2011ajh,HSP2006ajprp,DTO2010bmcg,F2010h}). Also, in the 
following we choose $8$ hub genes to give briefly their 
biological descriptions according to existing research.

\begin{table}[!]
\caption{Biological functions of specific hub genes playing important 
roles in hypertension.}
\label{genefunction}
\centering\scriptsize
\begin{tabular}[c]{llrl}
\hline
Gene                            & Official full name & Gene ID & Function\\
\hline
{\it Col4a1}                    & collagen, type IV, alpha 1 
& ${\;\,}$290${\,}$905 
& Related to epithelial cell differentiation \cite{TSS2009ccg,vABS2010hmg}\\
{\it Lcn2}                      & lipocalin 2 
& ${\;\,}$170${\,}$496 
& Related to cellular response to hydrogen peroxide \cite{OTC2011ceh}\\
{\it Cdk4}                      & cyclin-dependent kinase 4 
& ${\;\;\,\,}$94${\,}$201 
& Related to circadian rhythm; organ regeneration \cite{DEMT2001h}\\
{\it Fstl1}                     & follistatin-like 1 
& ${\;\;\,\,}$79${\,}$210 
& Related to aflatoxin B$_{1}$; ammonium chloride \cite{LWX2011art}\\
{\it Usp48}                     & ubiquitin specific peptidase 48 
& ${\;\,}$362${\,}$636 
& Related to ubiquitin-dependent protein catabolic process \cite{AVW2011c}\\
{\it Shc1}                      & SHC transforming protein 1 
& ${\;\;\,\,}$85${\,}$385 
& Related to activation of MAPK (mitogen-activated protein kinase) activity \\
                                &  &  & \cite{NYM2010cc}\\
{\it Nr1d2}                     & nuclear receptor subfamily 1, group D, 
& ${\;\,}$259${\,}$241 
& Related to regulation of transcription \cite{LBM2013gn}\\
                                & member 2 &  & \\
{\it Fzd2}                      & frizzled family receptor 2 
& ${\;\;\,\,}$64${\,}$512 
& Related to cellular response to growth factor stimulus \cite{GDA2002a,vGDS2002cr}\\
{\it Cdc2a} ({\it Cdk1})        & cyclin-dependent kinase 1 
& ${\;\;\,\,}$54${\,}$237 
& Related to cellular response to hydrogen peroxide and cell aging \cite{YQG2006bbrc}\\
{\it Hist1h2ai}                 & histone cluster 1, H2ai 
& ${\;\,}$502${\,}$129 
& Related to nucleosome assembly \cite{CWH2010jmcc}\\
{\it Ctsd}                      & cathepsin D 
& ${\;\,}$171${\,}$293 
& Related to autophagy; proteolysis; autophagic vacuole assembly \cite{T1992sjcli,TGK2009k}\\
{\it Casp6}                     & caspase 6 
& ${\;\;\,\,}$83${\,}$584 
& Related to acute inflammatory response to non-antigenic stimulus \cite{BZL2005p}\\
{\it Timp1}                     & TIMP metallopeptidase inhibitor 1 
& ${\;\,}$116${\,}$510 
& Related to aging; cartilage development; cell activation \cite{TNB2004ajh}\\
{\it Fbn1}                      & fibrillin 1 
& ${\;\;\,\,}$83${\,}$727 
& Related to kidney development; heart development \cite{SLW2011ajh,HSP2006ajprp}\\
{\it MCWA09\_96} ({\it Ppp3ca}) & protein phosphatase 3, catalytic subunit, 
& ${\;\;\,\,}$19${\,}$055 
& Related to cardiac hypertrophy \cite{DTO2010bmcg}\\
                                & alpha isoform &  & \\
{\it Sdc1}                      & syndecan 1 
& ${\;\;\,\,}$25${\,}$216 
& Related to inflammatory response \cite{F2010h}\\
\hline
\end{tabular}
\end{table}

The gene {\it Col4a1} encodes the major type \Rmnum{4} alpha collagen 
chain of basement membranes. It has been indicated that mutations 
in {\it Col4a1} result in a complex vascular phenotype encompassing 
defects in maintenance of vascular tone, endothelial cell function 
and blood pressure regulation \cite{vABS2010hmg}. Other studies 
have suggested that a SNP (single-nucleotide polymorphism) in the 
{\it Col4a1} gene is strongly associated with PWV (pulse wave 
velocity), an established independent predictor of adverse 
cardiovascular outcomes \cite{TSS2009ccg}.

The gene {\it Lcn2}, which has the most connections in the gene 
network, is a recently identified adipokine that belongs to the 
superfamily of lipocalins. It is recognized as a biomarker of obesity 
and inflammation, which are both risk factors for hypertension. Recent 
findings suggest that genetic variants in {\it Lcn2} may affect blood 
pressure \cite{OTC2011ceh}, and adiponectin has multiple protective 
effects on vascular endothelium. 

The gene {\it Cdk4} is a positive regulator of the cell cycle, which 
plays a role in cell cycle progression. Angiotensin \Rmnum{2}, an 
essential factor for hypertension, is an important modulator of cell 
growth through AT$_{1}$ receptors. Studies have reported when 
AT$_{1}$ receptors are stimulated {\it in vivo}, DNA synthesis is 
enhanced in blood vessels by activation of cyclin D1 and {\it Cdk4} 
\cite{DEMT2001h}.

The gene {\it Usp48} encodes a protein containing domains that 
associate it with the peptidase family C19. It is also known as the 
family 2 of ubiquitin carboxyl-terminal hydrolases, whose family 
members function as de-ubiquitinating enzymes, recognizing and 
hydrolyzing the peptide bond at the C-terminal glycine of ubiquitin. 
Recent data demonstrate that via the inhibition of {\it Usp48}, 
agonist activation of D$_{3}$R (dopamine D$_{3}$ receptor gene) 
promotes the degradation of NHE3 (sodium/hydrogen exchanger isoform 
3), and thus engenders natriuresis and regulates blood pressure 
\cite{AVW2011c}.

The gene {\it Fzd2} belongs to a class of highly conserved genes, 
which acts as cell--surface receptors for Wnt proteins. There is 
evidence that Wnt/Fzd signaling is involved in the formation and 
remodeling of the vasculature. Some existing data also show that 
{\it Fzd2} expressed in aortic smooth muscle cells is modulated by 
Ang \Rmnum{2}, and Ang \Rmnum{2} plays a crucial role in blood 
pressure regulation and cardiovascular homeostasis, both {\it in 
vitro} and {\it in vivo} \cite{GDA2002a,vGDS2002cr}.

The gene {\it Ctsd} encodes a lysosomal aspartyl protease. This 
proteinase has a variety of biological functions, such as degradating 
hemoglobin, serum albumin and myoglobin in endosome. It has been 
reported that genetic variation of the aspartic proteinases may have 
an effect on specific clinical diseases such as hypertension and 
ulcers \cite{T1992sjcli}. Other research suggests that {\it Ctsd} 
significantly positively correlates with the arterial hypertension 
stage as well as with histological grading of atherosclerotic lesions 
\cite{TGK2009k}.

The gene {\it Timp1} is linked to extracellular matrix fibrosis and 
is elevated in hypertension. Hypertension results in structural 
changes to the cardiac and vascular extracellular matrix (ECM). 
Matrix metalloproteinases (MMP), and their inhibitors (TIMP) may 
play a central role in the modulation of this matrix. Existing 
observations suggest a possible role for these surrogate markers 
of tissue ECM composition and the prognosis of cardiovascular 
events in hypertension \cite{TNB2004ajh}.

The gene {\it Fbn1} encodes a member of the fibrillin family. 
Recently, it is reported that {\it Fbn1} plays an important role in 
maintaining the physiological arterial stiffness of essential 
hypertension \cite{SLW2011ajh}. Meanwhile, {\it Fbn1} is a major 
component of the microfibrils that form a sheath surrounding the 
amorphous elastin. Defects in the {\it Fbn1} gene are associated 
with an increased risk of prevalent hypertension. The investigation 
also reveals that {\it Fbn1} may contribute to glomerular damage in 
hypertensive and diabetic kidney disease \cite{HSP2006ajprp}.

Since our network model is constructed by considering the 
correlations of GED, hub genes which have similar expression 
patterns with most of the genes with changed expression levels 
are typical representatives of the genes with strong correlation. 
From relevant research literature and the biological functions 
of these hub genes shown in Table \ref{genefunction}, we know 
that these hub genes are key (feature) genes that play important 
roles in hypertension. This shows that our construction of the 
gene network is suitable.

\section{Concluding remarks}
\label{sec:5}

Hypertension has the complexity and diversity of the genetic 
factors and pathogenesis. However, a comprehensive understanding 
of hypertension is still extremely deficient. In this study, 
based on microarray data, we attempt to reversely obtain the 
relationships between the hypertension-related genes, rebuild 
the structure of the gene network by visualization technology, 
and try to abstract the complex interactions between genes 
through calculating statistical characteristics of the gene 
network.

The network of hypertension-related genes is sparse, which has 
the following characteristics: 
(\rmnum{1}) The gene network shows a scale-free property with 
power-law degree distribution. A few nodes with large values 
of three centrality indices $\{C_{d},C_{b},C_{c}\}$ (or 
equivalently with large values of integrated centrality 
$C_{\mathrm{intgr}}$) correspond to hub genes; they are key 
(feature) genes involved in the formation of hypertension. 
(\rmnum{2}) The gene network has a small average path length 
($L=5.428$) and a large clustering coefficient ($C=0.6063$), 
i.e., the small-world property, indicating that the local 
disturbance to hubs would rapidly transfer to the whole 
network. This property reveals the direct influence of these 
hub genes on hypertension from another perspective, and 
implies possible novel molecular genetic signals. 
\linebreak[4](\rmnum{3}) The gene network of hypertension 
is a description of an abnormal state (illness), which 
exhibits assortative feature (assortativity coefficient 
$r=0.4065$) unlike most biological networks. The nodes with 
many connections tend to be connected to those with the 
similar type, showing that besides being affected by a few 
feature genes, hypertension is the consequence of accumulation 
of genetic changes and interaction caused by a variety of 
factors; this has been demonstrated by the existing research 
findings. Therefore, it is necessary to focus on further study 
of the interplay between genes in biological experiments.

In this paper, we construct the network model of 
hypertension-related genes in genetic level based on both 
biological knowledge and network theory. The hub genes 
({\it Col4a1}, {\it Lcn2}, {\it Cdk4}, etc.) in our network 
have been confirmed by biological/medical research to play 
important roles in hypertension. Furthermore, the network can 
also be analyzed based on actual functional correlation metrics 
with other threshold selection methods (e.g., 
\cite{CGBC2011itbe}) to get more biological information about 
hypertension in the future; it is believable that the results 
derived from those methods would be consistent and harmonic 
in comparison with the theoretical analysis of this paper, and 
they would complement each other. This study provides another 
perspective on expounding the molecular genetic mechanism, 
prevention, and individualization treatment of 
salt-sensitive hypertension. Meanwhile, the research may 
hopefully shed light on the development of network-based 
models (e.g., directed and/or weighted networks) of 
hypertension and other serious diseases, and explore the 
mutual regulatory relationships between genes of complex 
diseases, as well as contribute to finding new drug targets 
and developing novel ideas. At last, we expect that the 
complex network approach can provide an effective tool for 
analyzing the pathogenesis of critical illness.

\section*{Acknowledgments}

This work was supported in part by the National Natural Science 
Foundation of China (NSFC) (Grant Nos. 10565004 and 11365023) 
and the Specialized Research Fund for the Doctoral Program of 
Higher Education of China (SRFDP) (Grant No. 20050673001). Our 
special thanks go to the authors of Ref. \cite{LLW2008pg} for 
providing the online GED of the SS rat and two consomic rat 
strains, which are the basis of construction of our network 
model. The authors would like to thank Prof. S.-L. Peng and 
Dr. J.-J. Hao for their helpful discussions and suggestions, 
and the anonymous reviewer for calling our attention to Ref. 
\cite{CGBC2011itbe}.












\end{document}